\documentclass[aps,showpacs,showkeys,nofootinbib]{revtex4}
\usepackage{amsmath,amscd,amssymb}
\usepackage{color,epsfig}
\usepackage{xfrac}
\usepackage{latexsym}
\usepackage{mathrsfs}
\usepackage{graphicx}
\usepackage{bbm}      

\newcommand{\Vek}[1]{{\boldsymbol#1}}

\newcommand{\fract}[2]{{\textstyle\frac{#1}{#2}}}

\newcommand{\imu}{{\rm i}}

\usepackage{tikz}
\usetikzlibrary{decorations.markings}
\usetikzlibrary{decorations.pathreplacing}
\usetikzlibrary{arrows,decorations.pathmorphing,backgrounds,positioning,fit,petri,patterns}

\begin{document}

\baselineskip16pt

\title{Heavy Baryons with Strangeness in a Soliton Model}

\author{J. P. Blanckenberg, H. Weigel}

\affiliation{Physics Department, Stellenbosch University,
Matieland 7602, South Africa}

\begin{abstract}
We present results from a chiral soliton model calculation for the spectrum of 
baryons with a single heavy quark (charm or bottom) and non-zero strangeness. We 
treat the strange components within a three flavor collective coordinate 
quantization of the soliton that fully accounts for light flavor symmetry breaking. 
Heavy baryons emerge by binding a heavy meson to the soliton. The dynamics of this 
heavy meson is described by the heavy quark effective theory with finite mass 
effects included.
\end{abstract}



\maketitle

\section{Motivation}

Baryons containing heavy quarks such as charm or bottom form an excellent 
opportunity to study the binding of quarks to hadrons. Since there is no
exact solution to quantum-chromo-dynamics (QCD), various models and
approximations that focus on particular features of QCD are relevant. In the 
context of heavy baryons three are of particular importance. First, there is 
the heavy spin-flavor symmetry~\cite{Eichten:1980mw}\footnote{For reviews see 
Refs.~\cite{Neubert:1993mb}.} that governs the dynamics of heavy quarks. 
Second, there is the chiral symmetry that dictates the interactions among the 
light quarks. In addition to dynamical chiral symmetry breaking, there is 
substantial flavor symmetry breaking when the strange quark is involved. It 
is thus particularly interesting to investigate baryons that, in the valence 
quark picture, are composed of a single heavy quark and two  light, including 
strange, quarks.  Thirdly, generalizing QCD from three to arbitrarily many 
color degrees of freedom suggests to consider baryons as soliton configurations 
in an effective meson theory~\cite{Witten:1979kh}.

Our point of departure is a chiral soliton of meson fields built from up and 
down quarks~\cite{Adkins:1983ya}. States with good baryon quantum numbers 
are generated by quantizing the fluctuations about the soliton. The modes 
associated with (flavor) rotations have large non-harmonic components and 
consequently are treated as collective excitations. The Hamiltonian for these 
collective coordinates contains flavor symmetry breaking terms that slightly
suppress the non-harmonic contributions. The important feature is that this
Hamiltonian can be diagonalized exactly, {\it i.e.} the spectrum can be determined
beyond a perturbation expansion in the quark mass differences~\cite{Yabu:1987hm}. 
The resulting eigenvalues are associated with the strangeness contribution to the 
baryon masses. For the particular case of kaon-nucleon scattering this approach 
has been verified~\cite{Walliser:2005pi} to yield the correct resonance position.
Subsequently fields representing mesons with a single heavy quark are included. 
While their heavy quark components are subject to the heavy spin-flavor symmetry, 
their light ones couple to the light meson fields according to chiral symmetry such 
that the soliton generates an attractive potential for the heavy meson 
fields~\cite{Schechter:1995vr}. Combined with the soliton, a bound state in this 
potential builds the heavy baryon. (This is a generalization of the so-called bound 
state approach~\cite{Callan:1985hy} that, in the harmonic approximation, describes 
hyperons in the Skyrme model~\cite{Skyrme:1961vq}\footnote{While Ref.~\cite{Weigel:2008zz} 
comprehensively reviews soliton model studies, Ref.~\cite{Walliser:2005pi} thoroughly 
discusses the two above mentioned descriptions of strangeness in chiral soliton models, 
in particular with regard to the large $N_C$ limit.}.) The strangeness components of 
the heavy meson bound state are subject to the same collective coordinate treatment as 
the soliton explained above.

Shortly after the bound state approach in the Skyrme model of pseudoscalar mesons was 
applied to hyperons it was extended to heavier baryons~\cite{Rho:1992yy}. In those studies 
the relevance of the heavy spin-flavor symmetry was not yet recognized. Subsequently, also
heavy vector meson fields were included~\cite{Oh:1994zz}. More or less at the same time 
investigations were performed in the heavy limit scenario~\cite{Guralnik:1992dj,Harada:1997we}.
Those heavy limit studies neither included corrections to the heavy spin-flavor symmetry 
from finite masses nor strangeness degrees of freedom. In that case, baryons like $\Xi_c$ cannot 
be addressed. Strangeness was indeed included in Ref.~\cite{Momen:1993ax}, however, 
light flavor symmetry breaking was treated in a perturbation expansion and finite mass 
effects were omitted. This does not distinguish between even and odd parity or charm and 
bottom baryons and typically overestimates the binding energy of the heavy 
meson~\cite{Schechter:1995vr}. Also the parameters of the final energy formula were fitted 
rather than calculated from a realistic soliton model. These widespread bound state 
studies derive a potential for the meson fields from the soliton that is fixed in position. 
We note that this picture is strictly valid only in the large number of colors limit when 
the soliton is more massive than the heavy meson. Though this approach is a 
systematic and consistent expansion in the number of colors, kinematical corrections
should be expected in the real world with three colors.

Our soliton model calculation for the spectrum of heavy baryons will improve with regard 
to the following aspects: We take the parameters in the mass formula from an actual soliton 
model calculation (we allow for moderate adjustment of the light flavor symmetry breaking 
strength), go beyond the perturbation expansion in that symmetry breaking and construct 
the heavy meson bound state from a model that systematically incorporates finite mass 
corrections. Our model calculation will produce an extensive picture of baryons, from 
the nucleon up to the $\Omega_b$.  We will not consider doubly-heavy baryons, though.

The spectrum of heavy baryons has been investigated in other approaches as well.
A comprehensive account of the (non-relativistic) quark model approach is given 
in Ref.~\cite{Roberts:2007ni} with some newer results reported in 
Ref.~\cite{Lipkin:2011zz}. Relativistic effects are incorporated
within quark-diquark models~\cite{Ebert:2007nw}. QCD sum rules were not only used 
to obtain the spectrum~\cite{Zhang:2008pm}, but also to extract the heavy quark mass 
poles~\cite{Groote:1996em}. Lattice QCD calculations can be traced from 
Ref.~\cite{Brown:2014ena} that also studies baryons with more than one heavy quark. 
Finally, Ref.~\cite{Klempt:2009pi} contains comprehensive reviews on baryon spectroscopy 
that discuss a variety of approaches and may be consulted for further references.

\section{The Soliton Model}

In chiral Lagrangians the interaction terms are ordered by the number of 
derivatives acting on the pseudoscalar fields. The more derivatives there 
are, the more unknown parameters appear in the Lagrangian. Replacing these 
higher derivatives by resonance exchange term is advantageous because more 
information is available to determine the parameters. We thus consider a chiral 
soliton that is stabilized by vector mesons $\rho$ and $\omega$~\cite{Jain:1987sz} 
as a refinement of the Skyrme model~\cite{Skyrme:1961vq,Adkins:1983ya}.
Other shortcomings of the pseudoscalar soliton, like the neutron proton 
mass difference or the axial singlet matrix element of the nucleon 
are also solved when including light vector mesons~\cite{Weigel:2008zz}. 

The basic building block of the model is the chiral field 
$U={\rm exp}\left(\imu \sum_{a=1}^8\phi_a(x)\lambda_a/f_a\right)$, {\it i.e.}
the non-linear realization of the pseudoscalar octet field $\phi_a(x)$.  Here $f_a$ 
are the respective decay constants [$f_\pi=93{\rm MeV}$ (for $a=1,2,3$), 
$f_k=114{\rm MeV}$ (for $a=4,\ldots,7$). The case $a=8$ requires additional
input~\cite{Schechter:1992iz} but is not relevant here.] and
$\lambda_a$ are the eight Gell-Mann matrices of $SU(3)$. The static field configuration 
of the soliton is the hedgehog ansatz
\begin{equation}
U_0(\Vek{r})={\rm exp}\left[\Vek{\tau}\cdot\hat{\Vek{r}}F(r)\right]\,,\quad
\omega_\mu(\Vek{r})=\omega(r)\,g_{\mu0}\qquad {\rm and} \qquad
\rho^{(0)}_{im}(\Vek{r})=\epsilon_{ikm}\hat{r}_k\frac{G(r)}{r}\,.
\label{eq:cl1}
\end{equation}
The isovector $\Vek{\tau}=\left(\lambda_1,\lambda_2,\lambda_3\right)$ comprises the 
three Pauli matrices from the isospin subspace of flavor $SU(3)$. The spatial components 
of the $\omega_\mu$ and the time components of the $\Vek{\rho}_\mu$ fields are zero. 
For the latter, $i$ is an isospin/flavor index and $m=1,2,3$ labels its spatial 
components. The profile functions $F(r)$, $\omega(r)$ and $G(r)$ enter the classical 
energy functional, $E_{\rm cl}$. The profiles are determined by the minimization 
of $E_{\rm cl}$ , subject to boundary conditions that ensure unit baryon number:
\begin{equation}
F(0)=0\,,\quad \frac{d\omega_0(r)}{dr}\Big|_{r=0}=0
\quad {\rm and} \quad G(0)=-2\,.
\label{eq:bc1}
\end{equation}
All profile functions vanish asymptotically. Configurations that are suitable for 
quantization are obtained by introducing time dependent collective coordinates 
for the flavor orientation $A(t)\in SU(3)$
\begin{equation}
U(\Vek{r},t)=A(t) U_0(\Vek{r}) A^\dagger(t)
\quad {\rm and} \quad
\Vek{\tau}\cdot\Vek{\rho}_\mu(\Vek{r},t)=
A(t)\Vek{\tau}\cdot\Vek{\rho}^{(0)}_\mu(\Vek{r})A^\dagger(t)\,.
\label{eq:cc1}
\end{equation}
In addition profile functions are induced for the spatial components of $\omega_\mu$ 
and the time components of $\rho_\mu$~\cite{Meissner:1986js,Park:1991fb}. Defining 
eight angular velocities $\Omega_a$ via the time derivative of the collective coordinates
\begin{equation}
\frac{\imu}{2}\sum_{a=1}^8\Omega_a\lambda_a=
A^\dagger(t)\,\frac{dA(t)}{dt}\,,
\label{eq:cc2}
\end{equation}
allows a compact presentation of the Lagrange function for the collective 
coordinates from the light meson fields
\begin{equation}
L_l(\Omega_a)=-E_{\rm cl}+\frac{1}{2}\alpha^2\sum_{i=1}^3\Omega_i^2
+\frac{1}{2}\beta^2\sum_{\alpha=4}^7\Omega_\alpha^2
-\frac{\sqrt{3}}{2}\Omega_8\,.
\label{eq:cc3}
\end{equation}
It is obtained from the spatial integral over the Lagrange density with the
above described field configuration substituted. Note that the collective 
coordinates only appear via the angular velocities; $A$ does not appear explicitly. 
The last term, which is only linear in the time derivative, originates from the 
Wess-Zumino-Witten action~\cite{Wi83} that incorporates the QCD anomaly. The 
coefficients $\alpha^2$ and $\beta^2$ are moments of inertia for rotations in 
isospace\footnote{Because of the hedgehog structure it is equivalent to coordinate 
space.} and the strangeness subspace of flavor $SU(3)$, respectively. These moments 
of inertia  are functionals of profile functions and  the variational 
principle determines the induced components of the vector meson fields. 
The structure of the collective coordinate Lagrangian, Eq.~(\ref{eq:cc3}) is 
generic to all chiral models that support soliton solutions. The particular numerical
values for the classical energy and the moments of inertia are, of course, subject
to the particular model. Here we employ the calculation described in appendix 
A of Ref.~\cite{Park:1991fb} for the entries of Eq.~(\ref{eq:cc3}).

\section{Heavy Meson Bound State}

In effective meson theories, the heavy flavor enters via a heavy meson containing
a single heavy quark (charm or bottom) of mass $M$. The dynamics of the heavy meson 
follows the heavy flavor effective theory~\cite{Neubert:1993mb} 
that treats the pseudoscalar ($P$) and vector meson ($Q_\mu$) components equivalently. 
That is, in the limit $M\to\infty$ these components are part of a single multiplet
(The constant four--velocity $V^\mu$ characterizes the heavy quark rest frame.)
\begin{equation}
H=\frac{1}{2}\left(1+\gamma_\mu V^\mu\right)
\left(i\gamma_5P^\prime+\gamma^\mu Q^\prime_\mu\right)
\quad {\rm where} \quad
P^\prime={\rm e}^{-iMV\cdot x}P
\quad {\rm and} \quad
Q_\mu^\prime={\rm e}^{-iMV\cdot x}Q_\mu\,.
\label{eq:HH}
\end{equation}
The Lagrangian that describes the coupling of this multiplet to the light mesons
including the vector mesons $\rho$ and $\omega$ and respects the heavy spin-flavor symmetry 
is~\cite{Schechter:1992ue}
\begin{equation}
\frac{1}{M}\widetilde{\mathcal{L}}_H=
i V_\mu {\rm Tr}\left\{H D^\mu {\bar H}\right\}
- d {\rm Tr}\left\{H\gamma_\mu\gamma_5 p^\mu {\bar H}\right\}
- i \frac{\sqrt{2}c}{m_\rho}{\rm Tr}\left\{H\gamma_\mu\gamma_\nu 
F^{\mu\nu}(\rho){\bar H}\right\} + \ldots \ ,
\label{eq:hlimit}
\end{equation}
where ${\bar H}=\gamma_0H^\dagger\gamma_0$.
We take the covariant derivative to be\footnote{Symmetry allows to also include the light 
vector meson in this derivative at the expense of an unknown coupling constant. The bound 
state energies only show moderate sensitivity on that constant~\cite{Schechter:1995vr} 
so we omit it here.} $D_\mu=\partial_\mu+iv_\mu$. The chiral currents of the light pseudoscalar 
mesons are
$v_\mu,p_\mu=\frac{i}{2}\left(\sqrt{U}\partial_\mu\sqrt{U^\dagger}
\pm\sqrt{U^\dagger}\partial_\mu\sqrt{U}\right)$\,.
The heavy-light coupling constants $d\approx0.53$ and $c\approx1.60$ were determined 
from heavy meson decays. A field theory model that minimally extends to finite $M$ 
and $M^\ast$ for the pseudoscalar and vector components, respectively, has also been 
constructed in Ref.\cite{Schechter:1992ue} 
\begin{align}
\mathcal{L}_H&=\left(D_\mu P\right)^{\dagger}D^\mu P
-\frac{1}{2}\left(Q_{\mu\nu}\right)^{\dagger}Q^{\mu\nu}
-M^2P^{\dagger}P+M^{*2}Q_\mu^\dagger Q^\mu
\nonumber \\ &
+2iMd\left(P^\dagger p_\mu Q^\mu-Q^\dagger_\mu p^\mu P\right)
-\frac{d}{2}\epsilon^{\alpha\beta\mu\nu}
\left[\left(Q_{\nu\alpha}\right)^\dagger p_\mu Q_\beta+
Q^\dagger_\beta p_\mu Q_{\nu\alpha}\right]
\nonumber \\ &
-\frac{2\sqrt{2}icM}{m_V}\left\{
2Q^\dagger_\mu F^{\mu\nu}Q_\nu
-\frac{i}{M}\epsilon^{\alpha\beta\mu\nu}\left[
\left(D_\beta P\right)^\dagger F_{\mu\nu}Q_\alpha
+Q^\dagger_\alpha F_{\mu\nu}D_\beta P\right]\right\},
\label{eq:lagheavy}\end{align}
so that $\mathcal{L}_H\to\widetilde{\mathcal{L}}_H$ in the heavy limit.
Here $Q_{\mu\nu}$ and $F_{\mu\nu}$ are the field strength tensors of the heavy
and light vector mesons, respectively. The central feature is that, 
through the coupling to the light meson soliton, solutions for the heavy meson fields 
emerge with energy $0<\omega<M$, {\it i.e.} bound states. (Negative energy bound
states are also possible. Eventually they build pentaquark baryons that will 
not be considered here.) The most strongly bound solution has P-wave structure 
in the pseudoscalar component:
\begin{equation}
P=\frac{{\rm e}^{i\omega t}}{\sqrt{4\pi}}\Phi(r) 
\hat{\Vek{r}}\cdot \hat{\Vek{\tau}}\chi\, ,\qquad
Q_0=\frac{{\rm e}^{i\omega t}}{\sqrt{4\pi}}\Psi_0(r)\chi 
\quad {\rm and}\quad
Q_i=\frac{{\rm e}^{i\omega t}}{\sqrt{4\pi}}\left[i\Psi_1(r){\hat r}_i
+\fract{1}{2}\Psi_2(r)\epsilon_{ijk}{\hat r}_j\tau_k\right]\chi\,.
\label{eq:pansatz}
\end{equation}
Here $P$ and $Q_\mu$ are three component spinors whose flavor content is 
parameterized by the (constant) spinor $\chi$. Since the coupling to the light 
mesons occurs via a soliton in the isospin subspace, only the first two components 
of $\chi$ are non-zero. The four radial functions in Eq.~(\ref{eq:pansatz}) 
couple to the profiles of the static soliton, Eq.~(\ref{eq:cl1}) in linear 
differential equations. Normalizable solutions exist only for certain values of 
$\omega$. These are the bound wave-functions. Their construction, in particular 
with regard to finite $M$ corrections, and their normalization to carry unit heavy 
charge is explained in Refs.~\cite{Schechter:1995vr} and~\cite{Harada:1997we}, 
respectively. A heavy baryons is then a compound system of the soliton for the light 
flavors and the bound state of the heavy meson~\cite{Callan:1985hy}. There are also 
bound states in the $S$-wave channel in which the heavy meson field is parameterized as
(See Ref.~\cite{Oh:1994zz} for parameterizations of higher angular momenta.)
\begin{equation}
P=\frac{{\rm e}^{i\omega t}}{\sqrt{4\pi}}\Phi(r) \chi\, ,\qquad
Q_0=\frac{{\rm e}^{i\omega t}}{\sqrt{4\pi}}\Psi_0(r) \hat{\Vek{r}}\cdot \hat{\Vek{\tau}}\chi 
\quad {\rm and}\quad
Q_i=\frac{{\rm e}^{i\omega t}}{\sqrt{4\pi}}\left[
\Psi_1(r){\hat r}_i\hat{\Vek{r}}\cdot \hat{\Vek{\tau}}
+\Psi_2(r)r \Vek{\tau}\cdot\partial_i\hat{\Vek{r}}\right]\chi\,.
\label{eq:sansatz}
\end{equation}
They combine with the soliton to form negative parity heavy 
baryons~\cite{Schechter:1995vr,Harada:1997we}. For convenience we have used equal 
symbols for the $S$ and $P$--wave profile functions but, of course, they are different.
The computation of the bound state energies $\omega$ from identifying localized solutions 
to the equations of motions that arise by substituting the parameterizations, 
Eqs.~(\ref{eq:pansatz}) and~(\ref{eq:sansatz}), into the Euler-Lagrange equations of 
Eq.~(\ref{eq:lagheavy}) is detailed in appendix A of Ref.~\cite{Schechter:1995vr}.
That reference also provides figures of the resulting profile functions.

The heavy meson fields must also account for the collective flavor rotation introduced 
in Eq.~(\ref{eq:cc1}). This enforces the substitution
\begin{equation}
P\,\longrightarrow\, A(t) P \qquad {\rm and}\qquad
Q_\mu\,\longrightarrow\, A(t) Q_\mu\,,
\label{eq:hcc1}
\end{equation}
where the right hand sides contain the fields introduced in Eq.~(\ref{eq:pansatz}). 
This gives non--zero strange components of the heavy mesons and couples the heavy meson 
strange quark to that of the soliton. Substituting this flavor rotating configuration 
into the Lagrange density and integrating over space provides the collective coordinate Lagrange 
function from the heavy fields 
\begin{equation}
L_h(\Omega_a)=-\omega\chi^\dagger \chi 
+\frac{1}{2\sqrt{3}}\chi^\dagger\Omega_8 \chi
+\rho\chi^\dagger\left(\Vek{\Omega}\cdot\frac{\Vek{\tau}}{2}\right)\chi\,.
\label{eq:hcc2}
\end{equation}
Again, the flavor rotation matrix $A$ does not appear explicitly. With the time 
dependence of the collective coordinates, terms that involve 
$\sum_{a=1}^8\lambda_a\Omega_a$ enter. In the heavy meson sector the quadratic 
terms provide the bound state contributions to the moments of inertia $\alpha^2$ 
and~$\beta^2$. Since the bound state wave-functions are strongly localized 
around the center of the soliton\footnote{Their asymptotic behavior is 
${\rm e}^{-|\omega|r}\sim {\rm e}^{-Mr}$ compared to ${\rm e}^{-m_\pi r}$ of
the chiral field.} the latter dominates the moments of inertia. It is thus safe
to only retain the linear terms in Eq.~(\ref{eq:hcc2}). At that order only $a=1,2,3$ 
and $a=8$ survive because the bound states do not have any strangeness components. 
The normalization of the bound state wave-function dictates the coefficients in the
first and second terms. The hyperfine splitting parameter $\rho$ is a functional of 
all profile functions, including some of the induced light vector fields. Its explicit 
expression is given in Eqs.~(B.1)-(B.4) of Ref.~\cite{Harada:1997we}, 
where it is called $\chi_P$ and $\chi_S$ for $P$- and $S$-wave channels, respectively.

\section{Quantization in SU(3), Symmetry Breaking and Hyperfine Splitting}

Before we construct a Hamilton operator for the collective coordinates 
via Legendre transformation of the Lagrangian $L_l+L_h$ we recall that
the rotations introduced in Eq.~(\ref{eq:cc1}) are not exact zero modes in any
sensitive model. The reason is that $SU(3)$ flavor symmetry is explicitly broken
by different (current) quark masses. This breaking is measured by the ratio
\begin{equation}
x=\frac{2m_s}{m_u+m_d}\,,
\label{eq:sbr1}
\end{equation}
where the $m_q$ are the current quark masses of the respective quarks. It can be 
estimated from meson data~\cite{Gasser:1982ap,Harada:1995sj,Schechter:1992iz}.
In early soliton model studies this ratio was considered to be quite large, 
$x\approx 30$~\cite{Schechter:1992iz}, or even bigger~\cite{Jain:1989kn}. This was 
accompanied by sizable symmetry breaking among the hyperons~\cite{Park:1991fb}. 
Later this ratio was re-evaluated and found to be somewhat smaller: 
$20\le x\le25$~\cite{Harada:1995sj}. 
Thus it is appropriate to consider this ratio for the (light) flavor symmetry 
breaking as a tunable parameter. Then symmetry breaking adds to the collective 
coordinate Lagrangian
\begin{equation}
L_{\rm sb}(A)=-\frac{x}{2}\widetilde{\gamma}\left[1-D_{88}(A)\right]\,,
\label{eq:sbr2}
\end{equation}
where $D_{ab}=\fract{1}{2}{\rm tr}\left[\lambda_aA\lambda_bA^\dagger\right]$
parameterizes the adjoint representation of the collective rotations. The 
coefficient~$\widetilde{\gamma}$ is again a functional of the profile functions 
and acquires its main contribution from the classical fields, Eq.~(\ref{eq:cl1}). It 
can be computed in any soliton model. (In the literature $\gamma=x\widetilde{\gamma}$ 
is typically used.) The heavy mesons also contribute to the symmetry breaking
parameter by appropriately substituting mass matrices in Eq.~(\ref{eq:lagheavy}).
For example, for the charm heavy meson in the $P$--wave channel we have
\begin{equation}
\gamma=\gamma_{\rm soliton}+\int dr r^2\left[
\left(m_D^2-m_{D_s}^2\right)\Phi^2+
\left(m_{D^*}^2-m_{D^*_s}^2\right)
\left(-\Psi_0^2+\Psi_1^2+\frac{1}{2}\Psi_2^2\right)\right]\,.
\label{eq:gammaheavy}
\end{equation}
Numerically this contribution is small and can easily be compensated by a slight 
change of $x$.

We have now collected all terms for the collective coordinate Lagrangian
$L(A,\Omega)=L_l(\Omega)+L_h(\Omega)+L_{\rm sb}(A)$ and can construct the 
Hamilton operator by Legendre transformation,
\begin{align}
H(A,R_a,\chi)&=E_{\rm cl}+\frac{1}{2}
\left(\frac{1}{\alpha^2}-\frac{1}{\beta^2}\right)\sum_{i=1}^3R_i^2
+\frac{1}{2\beta^2}\sum_{a=1}^8R_a^2
+\frac{x}{2}\widetilde{\gamma}\left[1-D_{88}(A)\right]
\cr\cr & \hspace{2cm}
-\frac{3}{8\beta^2}\left(1-\frac{1}{3}\chi^\dagger\chi\right)^2
+|\omega|\chi^\dagger\chi+H_{\rm hf}\,,
\label{eq:ham1}
\end{align}
where $R_a=\frac{\partial L}{\partial \Omega_a}$ defines the said Legendre 
transformation. The $R_a$ are the right generators of $SU(3)$ since
$\left[A,R_a\right]=A(\lambda_a/2)$ upon canonical quantization.
The spinors $\chi$ contain annihilation 
and creation operators for the heavy meson bound state. They are quantized as 
ordinary harmonic oscillators. In particular $\chi^\dagger\chi$ is the number 
operator for the heavy meson bound state. Since we are considering hadrons with 
a single heavy quark, contributions that are quartic in $\chi$ have been omitted 
for consistency. (In the square a term that is explicitly of quartic order is maintained 
because it cancels a similar term in $\sum_{a}R_a^2$, {\it cf.} subsection 
below.) The hyperfine splitting part, $H_{\rm hf}$, that emerges from the last 
term in Eq.~(\ref{eq:hcc2}), will be discussed later.

\subsection{SU(3) diagonalization}

The Hamiltonian, Eq.~(\ref{eq:ham1}) is not complete without the constraint 
\begin{equation}
Y_R=\fract{2}{\sqrt{3}}R_8=1-\fract{1}{3}\chi^\dagger\chi
\label{eq:constraint}\,,
\end{equation}
that arises from the terms linear in $\Omega_8$ in Eqs.~(\ref{eq:cc3}) 
and~(\ref{eq:hcc2}). Thus the heavy baryons have right hypercharge~$\sfrac{2}{3}$. 
Since the zero strangeness components of any $SU(3)$ representation has equal 
hypercharge and right hypercharge the $SU(3)$ coordinates must be quantized 
as diquarks for heavy baryons~\cite{Momen:1993ax}. The most relevant diquark 
representations are the antisymmetric anti-triplet and the symmetric sextet.

When symmetry breaking is included, elements of higher dimensional representations 
with the same flavor and $R_{1,2,3}$ quantum numbers are admixed. We first determine 
the quantum number $r$ in the intrinsic spin $\sum_{i=1}^3R_i^2=r(r+1)$: In 
addition to its dimensionality, an $SU(3)$ representation is characterized by two 
sets of quantum numbers $(I,I_3,Y)$ for the flavor and $(r,r_3,Y_R)$ for the $R_a$ 
degrees of freedom, respectively.  The flavor generators are $L_a=\sum_{b=1}^8D_{ab}R_b$ 
with $L_{1,2,3}=I_{1,2,3}$ and $Y=\fract{2}{\sqrt{3}}L_8$ being the observables. 
Low-dimensional representations (such as the anti-triplet and the sextet) are non-degenerate 
and their elements with $Y=Y_R$ have $\Vek{I}=\Vek{R}$. Thus $r$ equals the 
isospin ($I$) of the zero strangeness element within an $SU(3)$ representation: 
the anti-triplet has $r=0$ and the sextet has $r=1$. Symmetric and antisymmetric $SU(3)$ 
representations do not mix under symmetry breaking. Hence $r=0$ and $r=1$ for a heavy baryon 
whose diquark component builds up from the anti-triplet and sextet, respectively. 
The admixture of higher dimensional representations has been estimated in a perturbation 
expansion for hyperons~\cite{Park:1989wz} and heavy baryons~\cite{Momen:1993ax}. 
It can also be done exactly within the so-called Yabu-Ando approach~\cite{Yabu:1987hm}. 
The starting point is an Euler angle representation of the collective coordinates $A$ in 
which the conjugate momenta $R_a$ are differential operators. Then the eigenvalue equation
\begin{equation}
\left\{\sum_{a=1}^8R_a^2
+(x\widetilde{\gamma}\beta^2)\left[1-D_{88}(A)\right]\right\}\Psi(A)
=\epsilon\Psi(A)
\label{eq:ham2}
\end{equation}
is cast into a set of coupled ordinary second order differential equations. 
The single variable is the strangeness changing angle in $A$. The particular 
setting of the differential equations depends on the considered flavor quantum 
numbers. For ordinary baryons ($Y_R=1$) this treatment is reviewed in 
Ref.~\cite{Weigel:2008zz} and the results for diquark wave-functions that enter the 
heavy baryon wave-functions ($Y_R=\sfrac{2}{3}$) are reported in 
Ref.~\cite{Blanckenberg:2014}. Having obtained the $SU(3)$-flavor eigenvalue 
$\epsilon$ from the differential equations we simplify the $SU(3)$
part and write
\begin{equation}
H(A,R_a,\chi)\,\longrightarrow\,
H(\chi)=E_{\rm cl}+ 
\left(\frac{1}{\alpha^2}-\frac{1}{\beta^2}\right)\frac{r(r+1)}{2}
+\frac{\epsilon}{2\beta^2}
-\frac{3}{8\beta^2}\left(1-\frac{1}{3}\chi^\dagger\chi\right)^2
+|\omega|\chi^\dagger\chi+H_{\rm hf}\,.
\label{eq:ham3}
\end{equation}
The dependence of the eigenvalues $\epsilon$ on $x$ varies with spin and 
isospin. Hence there is implicit hyperfine splitting, however, it also appears 
explicitly as we discuss next.

\subsection{Hyperfine splitting}

The eigenstates of the Hamiltonian, Eq.~(\ref{eq:ham1}) are combinations in which 
each term is a product of two factors, one is a function of $A$ and the other of~$\chi$. 
The combinations are such that eigenstates of flavor and total spin are generated.
The flavor information is completely contained in $A$ because flavor transformations
correspond to multiplying $A$ by unitary matrices from the left. To construct
total spin eigenstates we consider the effect of spatial rotations. The soliton 
is the hedgehog configuration and spatial rotations are equivalent
to multiplying $A$ by unitary $SU(2)$ matrices from the right. For the heavy
meson bound state this multiplication must be compensated by an additional 
flavor transformation of the spinor $\chi$. Thus the total spin is
\begin{equation}
\Vek{J}=-\Vek{R}-\chi^\dagger\frac{\Vek{\tau}}{2}\chi\,.
\label{eq:totJ}
\end{equation}
Calling $j$ the spin of the considered baryon this implies 
$\Vek{R}\cdot\langle \Vek{\tau}\rangle= j(j+1)-r(r+1)-\frac{3}{4}\sim j(j+1)-r(r+1)$, 
where the expectation value refers to the heavy meson bound state. In the approximation 
we have again omitted terms that formally are quartic in $\chi$.  This scalar product 
appears in the Legendre transformation with respect to $\Vek{\Omega}$,
\begin{equation}
\frac{\partial L}{\partial \Vek{\Omega}}\cdot\Vek{\Omega}
-\frac{1}{2\alpha^2}\Vek{\Omega}^2
-\rho\chi^\dagger \left( \Vek{\Omega}\cdot\frac{\Vek{\tau}}{2}\right) \chi
=\frac{1}{2\alpha^2}\Vek{R}^2+\frac{\rho}{\alpha^2}\Vek{R}\cdot
\chi^\dagger\frac{\Vek{\tau}}{2}\chi\,.
\label{eq:hf1}
\end{equation}
Collecting pieces we get the mass formula
\begin{equation}
M=\left(\frac{1}{\alpha^2}-\frac{1}{\beta^2}\right)\frac{r(r+1)}{2}
+\frac{\epsilon}{2\beta^2}
-\frac{3}{8\beta^2}\left(1-\frac{N}{3}\right)^2
+|\omega|N+\frac{\rho}{2\alpha^2}\left[j(j+1)-r(r+1)\right]N\,,
\label{eq:master}
\end{equation}
where $N=0,1$ counts the number of heavy valence quarks contained in the
considered baryon. It has been included in the hyperfine splitting term since 
ordinary baryons have $r=j$. We have collected the leading contributions 
to the baryon energy in the large number of colors ($N_C$) expansion. 
However, a contribution $\mathcal{O}(N_C^0)$ is missing, the vacuum polarization 
energy $E_{\rm vac}$. It is the quantum correction to the classical energy $E_{\rm cl}$
that cannot be rigorously computed because the theory is not renormalizable. 
Estimates in the Skyrme model suggest that $E_{\rm vac}$ considerably reduces
$E_{\rm cl}$~\cite{Meier:1996ng}. We circumvent this limitation by only 
considering mass difference for which $E_{\rm cl}$ and $E_{\rm vac}$
cancel and consequently omit these terms from Eq.~(\ref{eq:master}). 

This quantization scheme predicts two heavy $\Xi$ baryons with spin $j=\sfrac{1}{2}$: 
one has $r=0$ and the other $r=1$. In an $SU(3)$ symmetric world the former 
would be an anti-triplet state and the latter a sextet state. There is no mixing
between these baryons because $\left[H,\Vek{R}^2\right]=0$. For $j=\sfrac{3}{2}$
only one heavy $\Xi$ baryon emerges in this scheme since then $r=1$ is required.
For the $\Xi$ hyperon there is also only a single option with $j=\sfrac{1}{2}$
that is build from the octet state. This counting suggests to relate $r$ to the 
intermediate spin $J_m$ defined in Ref.~\cite{Rho:1992yy}.

\section{Numerical Results}

As mentioned above, we consider mass differences, because the model predictions
for the absolute masses are subject to uncontrollable quantum contributions.

We find the energy eigenvalue $\epsilon$ in Eq.~(\ref{eq:ham2}) for all baryons 
and then compute their energies according to Eq.~(\ref{eq:master}).
We adopt the $SU(3)$ parameters from Ref.~\cite{Park:1991fb}:
$\alpha^2=5.144/{\rm GeV}$, $\beta^2=4.302/{\rm GeV}$
and\footnote{In the notation of Ref.~\cite{Park:1991fb}
$\widetilde{\gamma}=(\delta^\prime/\delta^{\prime\prime})\gamma$.}
$\widetilde{\gamma}=47{\rm MeV}$. For the heavy sector, the same soliton model
was used in Ref.~\cite{Harada:1997we} to compute the bound state energies $\omega$
and hyperfine parameters $\rho$ for both the $P$- and $S$-wave channels. From the model 
calculation described in section III the following bound state parameters were obtained
\begin{equation}
\omega_P=1326{\rm MeV}\,, \quad \rho_P=0.140 \,, \quad
\omega_S=1572{\rm MeV}\,, \quad \rho_S=0.181 
\label{eq:Cpara}
\end{equation}
and
\begin{equation}
\omega_P=4494{\rm MeV}\,, \quad \rho_P=0.053 \,, \quad
\omega_S=4663{\rm MeV}\,, \quad \rho_S=0.046 
\label{eq:Bpara}
\end{equation}
in the charm and bottom sector, respectively (Ref.~\cite{Harada:1997we} lists
the binding energies $\omega_{P,S}-M_D$ and $\omega_{P,S}-M_B$.). Then we are left 
with a single parameter, the effective symmetry breaking $x$ defined
in Eq.~(\ref{eq:sbr1}), that is not fully determined. We list our results for 
the charm and bottom baryon spectra in table \ref{tab:results}, that also 
contains the data for experimentally observed candidates~\cite{Agashe:2014kda}. 
We note that most of the quantum numbers listed in Ref.~\cite{Agashe:2014kda} are 
adapted from the quark model and stress that $r$ is not a physical observable. Hence 
assigning the experimental results for $\Xi$ type baryons to a particular $r$ value 
is a prediction.  Ref.~\cite{Agashe:2014kda} furthermore lists
$\Lambda_c(2625)$ and $\Lambda_b(5920)$ with spin $j=\sfrac{3}{2}$ that are not
contained in our approach: We require $|j-r|=\sfrac{1}{2}$ but the $\Lambda$'s
have neither strangeness nor isospin so they must have $r=0$ and $j=\sfrac{1}{2}$.
We complete the picture by including the corresponding results for the
low-lying non-heavy baryons in table~\ref{tab:light}.

\begin{table}
\centerline{\begin{tabular}{c|c||c|c|c|c||c|c|c|c||c|c|c|c}
&&\multicolumn{4}{c||}{$x=25$} &
\multicolumn{4}{c||}{$x=30$} &
\multicolumn{4}{c}{expt.~\cite{Agashe:2014kda}}
\cr
\hline
$(I,j,r)$ & &\multicolumn{2}{|c}{pos. par.}&
\multicolumn{2}{|c||}{neg. par.}&
\multicolumn{2}{|c}{pos. par.}&
\multicolumn{2}{|c||}{neg. par.}&
\multicolumn{2}{|c}{pos. par.}&
\multicolumn{2}{|c}{neg. par.}\cr
\hline
&  & $\Delta_N$ & $\Delta_c$
& $\Delta_N$ & $\Delta_c$
& $\Delta_N$ & $\Delta_c$
& $\Delta_N$ & $\Delta_c$
& $\Delta_N$ & $\Delta_c$
& $\Delta_N$ & $\Delta_c$
\cr \hline
$(0,\sfrac{1}{2},0)$ & $\Lambda_c$ 
& 1230 & 0 & 1479 & 249 & 1233 & 0 & 1482 & 249 &
1347 & 0 & 1653 & 306 \cr
$(1,\sfrac{1}{2},1)$ &$\Sigma_c$ 
& 1423 & 193 & 1664 & 434 & 1425 & 192 & 1666 & 433 &
1515 & 168 & -- & -- \cr
$(\sfrac{1}{2},\sfrac{1}{2},0)$ & $\Xi_c$ 
& 1446 & 216 & 1695 & 465 & 1486 & 253 & 1735 & 502 &
1529 & 186 & 1851 & 504 \cr
$(0,\sfrac{1}{2},1)$ &$\Omega_c$ 
& 1693 & 463 & 1934 & 704 & 1756 & 523 & 1997 & 764 &
1756 & 409 & -- & -- \cr
$(\sfrac{1}{2},\sfrac{1}{2},1)$ & $\Xi_c$ 
& 1557 & 328 & 1798 & 569 & 1588 & 355 & 1829 & 596 &
1637 & 290 & -- & -- \cr
$(1,\sfrac{3}{2},1)$ & $\Sigma_c$ 
& 1464 & 234 & 1717 & 487 & 1466 & 233 & 1719 & 486 &
1579 & 232 & -- & -- \cr
$(\sfrac{1}{2},\sfrac{3}{2},1)$ & $\Xi_c$ 
& 1598 & 369 & 1851 & 622 & 1629 & 396 & 1882 & 649 &
1706 & 359 & 1876 & 529(?) \cr
$(0,\sfrac{3}{2},1)$ & $\Omega_c$ 
& 1734 & 504 & 1987 & 757 & 1797 & 564 & 2050 & 817&
1831 & 484 & -- & -- \cr
\hline\hline
\end{tabular}}
\centerline{\begin{tabular}{c|c||c|c|c|c||c|c|c|c||c|c|c|c}
&  & $\Delta_N$ & $\Delta_b$
& $\Delta_N$ & $\Delta_b$
& $\Delta_N$ & $\Delta_b$
& $\Delta_N$ & $\Delta_b$
& $\Delta_N$ & $\Delta_b$
& $\Delta_N$ & $\Delta_b$
\cr \hline
$(0,\sfrac{1}{2},0)$ & $\Lambda_b$ 
& 4391 & 0 & 4560 & 168 & 4394 & 0 & 4563 & 168 &
4681 & 0 & 4973 & 292 \cr
$(1,\sfrac{1}{2},1)$ & $\Sigma_b$ 
& 4601 & 210 & 4771 & 380 & 4603 & 209 & 4773 & 379 &
4872 & 191 & -- & -- \cr
$(\sfrac{1}{2},\sfrac{1}{2},0)$ & $\Xi_b$ 
& 4608 & 216 & 4776 & 385 & 4647 & 253 & 4816 & 421 &
4855 & 174 & -- & -- \cr
$(0,\sfrac{1}{2},1)$ & $\Omega_b$ 
& 4871 & 480 & 5041 & 650 & 4935 & 540 & 5105 & 710 &
5110 & 429 & -- & -- \cr
$(\sfrac{1}{2},\sfrac{1}{2},1)$ & $\Xi_b$ 
& 4736 & 345 & 4906 & 514 & 4766 & 372 & 4936 & 542 &
-- & -- & -- & -- \cr
$(1,\sfrac{3}{2},1)$ & $\Sigma_b$ 
& 4617 & 226 & 4785 & 393 & 4619 & 225 & 4787 & 392 &
4983 & 212 & -- & -- \cr
$(\sfrac{1}{2},\sfrac{3}{2},1)$ & $\Xi_b$ 
& 4751 & 360 & 4919 & 528 & 4782 & 387 & 4950 & 555 &
5006 & 325 & -- & -- \cr
$(0,\sfrac{3}{2},1)$ & $\Omega_b$ 
& 4887 & 496 & 5055 & 664 & 4950 & 556 & 5118 & 724 &
-- & -- & -- &\hspace{0.34cm}--\hspace{0.34cm}
\end{tabular}}
\caption{\label{tab:results}Model results for the mass differences of the charm 
and bottom baryons: $\Delta_N=M-M_N$, $\Delta_c=M-M_{\Lambda_c}$ and
$\Delta_b=M-M_{\Lambda_b}$ with the $M$'s computed from Eq.~(\ref{eq:master})
in comparison with available experimental data. The spin and isospin of a
considered baryon are $I$ and $j$. The $SU(3)$ quantum number $r$ is defined in 
the text. All data are in MeV. See text for explanation of question mark on $\Xi_c$.}
\end{table}
\begin{table}
\begin{tabular}{c|c|c|c|c|c|c|c||c}
& $\Lambda$ & $\Sigma$ & $\Xi$ & $\Delta$ & $\Sigma^\ast$ 
&$\Xi^\ast$ & $\Omega$ & $\Xi-\Sigma$ \cr\hline
$x=25$ & 134 & 218 & 320 & 324 & 438 & 551 & 661 & 101 \cr
$x=30$ & 162 & 253 & 404 & 323 & 461 & 601 & 740 & 151 \cr \hline
expt.  & 177 & 254 & 379 & 293 & 446 & 591 & 733 & 125
\end{tabular}
\caption{\label{tab:light}Mass differences for 
non-heavy baryons with respect to the nucleon in MeV.}
\end{table}

\begin{figure}[hb]
\centering
\begin{tikzpicture}[scale=6.3][c]
\clip (-0.125,0.95) rectangle (2.6,2.03);
\tikzstyle{slantedb}=[pattern=north west lines, pattern color=blue]

\def\width{0.135}
\def\yzero{1}
\def\Boffset{1.68}
\def\Bscale{0.67}

\draw[black] (0,\yzero) -- (0,2.010);
\draw[black] (1*\width,\yzero) -- (1*\width,2.010);
\draw[black] (2*\width,\yzero) -- (2*\width,2.0100);
\draw[black] (3*\width,\yzero) -- (3*\width,2.0100);
\draw[black] (4*\width,\yzero) -- (4*\width,2.0100);
\draw[black] (5*\width,\yzero) -- (5*\width,2.0100);
\draw[black] (6*\width,\yzero) -- (6*\width,2.0100);
\draw[black] (7*\width,\yzero) -- (7*\width,2.0100);
\draw[black] (8*\width,\yzero) -- (8*\width,2.0100);
\draw[black] (11*\width,\yzero) -- (11*\width,2.0100);
\draw[black] (12*\width,\yzero) -- (12*\width,2.0100);
\draw[black] (13*\width,\yzero) -- (13*\width,2.0100);
\draw[black] (14*\width,\yzero) -- (14*\width,2.0100);
\draw[black] (15*\width,\yzero) -- (15*\width,2.0100);
\draw[black] (16*\width,\yzero) -- (16*\width,2.0100);
\draw[black] (17*\width,\yzero) -- (17*\width,2.0100);
\draw[black] (18*\width,\yzero) -- (18*\width,2.0100);
\draw[black] (19*\width,\yzero) -- (19*\width,2.0100);

\draw[black] (0*\width,1.347) -- (1*\width,1.347);
\draw[black] (1*\width,1.515) -- (2*\width,1.515);
\draw[black] (2*\width,1.529) -- (3*\width,1.529);
\draw[black] (3*\width,1.756) -- (4*\width,1.756);
\draw[black] (4*\width,1.637) -- (5*\width,1.637);
\draw[black] (5*\width,1.579) -- (6*\width,1.579);
\draw[black] (6*\width,1.706) -- (7*\width,1.706);
\draw[black] (7*\width,1.831) -- (8*\width,1.831);

\node (oh) at (0.5*\width,0.02+1.347) {1347};
\node (oh) at (1.5*\width,0.02+1.515) {1515};
\node (oh) at (2.5*\width,0.02+1.529) {1529};
\node (oh) at (3.5*\width,0.02+1.756) {1756};
\node (oh) at (4.5*\width,0.02+1.637) {1637};
\node (oh) at (5.5*\width,0.02+1.579) {1579};
\node (oh) at (6.5*\width,0.02+1.706) {1706};
\node (oh) at (7.5*\width,0.02+1.831) {1831};

\draw[black] (11*\width,4.681*\Bscale-\Boffset) -- (12*\width,4.681*\Bscale-\Boffset);
\draw[black] (12*\width,4.872*\Bscale-\Boffset) -- (13*\width,4.872*\Bscale-\Boffset);
\draw[black] (13*\width,4.855*\Bscale-\Boffset) -- (14*\width,4.855*\Bscale-\Boffset);
\draw[black] (14*\width,5.110*\Bscale-\Boffset) -- (15*\width,5.110*\Bscale-\Boffset);
\draw[black] (16*\width,4.893*\Bscale-\Boffset) -- (17*\width,4.893*\Bscale-\Boffset);
\draw[black] (17*\width,5.006*\Bscale-\Boffset) -- (18*\width,5.006*\Bscale-\Boffset);

\node (oh) at (11.5*\width,0.02+4.681*\Bscale-\Boffset) {4681};
\node (oh) at (12.5*\width,0.02+4.872*\Bscale-\Boffset) {4872};
\node (oh) at (13.5*\width,0.02+4.855*\Bscale-\Boffset) {4855};
\node (oh) at (14.5*\width,0.02+5.110*\Bscale-\Boffset) {5110};
\node (oh) at (16.5*\width,0.02+4.893*\Bscale-\Boffset) {4893};
\node (oh) at (17.5*\width,0.02+5.006*\Bscale-\Boffset) {5006};

\draw[slantedb] (0*\width,1.230) rectangle (1*\width,1.233);
\draw[slantedb] (1*\width,1.423) rectangle (2*\width,1.425);
\draw[slantedb] (2*\width,1.446) rectangle (3*\width,1.486);
\draw[slantedb] (3*\width,1.693) rectangle (4*\width,1.756);
\draw[slantedb] (4*\width,1.557) rectangle (5*\width,1.588);
\draw[slantedb] (5*\width,1.464) rectangle (6*\width,1.466);
\draw[slantedb] (6*\width,1.598) rectangle (7*\width,1.629);
\draw[slantedb] (7*\width,1.734) rectangle (8*\width,1.797);

\draw[slantedb] (11*\width,4.391*\Bscale-\Boffset) rectangle (12*\width,4.394*\Bscale-\Boffset);
\draw[slantedb] (12*\width,4.601*\Bscale-\Boffset) rectangle (13*\width,4.603*\Bscale-\Boffset);
\draw[slantedb] (13*\width,4.608*\Bscale-\Boffset) rectangle (14*\width,4.647*\Bscale-\Boffset);
\draw[slantedb] (14*\width,4.871*\Bscale-\Boffset) rectangle (15*\width,4.935*\Bscale-\Boffset);
\draw[slantedb] (15*\width,4.736*\Bscale-\Boffset) rectangle (16*\width,4.766*\Bscale-\Boffset);
\draw[slantedb] (16*\width,4.617*\Bscale-\Boffset) rectangle (17*\width,4.619*\Bscale-\Boffset);
\draw[slantedb] (17*\width,4.751*\Bscale-\Boffset) rectangle (18*\width,4.782*\Bscale-\Boffset);
\draw[slantedb] (18*\width,4.887*\Bscale-\Boffset) rectangle (19*\width,4.950*\Bscale-\Boffset);

\node (oh) at (0.5*\width,\yzero-0.02) {$\Lambda_c$};
\node (oh) at (1.5*\width,\yzero-0.02) {$\Sigma_c$};
\node (oh) at (2.5*\width,\yzero-0.02) {$\Xi_c$};
\node (oh) at (3.5*\width,\yzero-0.02) {$\Omega_c$};
\node (oh) at (4.5*\width,\yzero-0.02) {$\Xi_c$};
\node (oh) at (5.5*\width,\yzero-0.02) {$\Sigma_c^\ast$};
\node (oh) at (6.5*\width,\yzero-0.02) {$\Xi_c^\ast$};
\node (oh) at (7.5*\width,\yzero-0.02) {$\Omega_c^\ast$};

\node (oh) at (11.5*\width,\yzero-0.02) {$\Lambda_b$};
\node (oh) at (12.5*\width,\yzero-0.02) {$\Sigma_b$};
\node (oh) at (13.5*\width,\yzero-0.02) {$\Xi_b$};
\node (oh) at (14.5*\width,\yzero-0.02) {$\Omega_b$};
\node (oh) at (15.5*\width,\yzero-0.02) {$\Xi_b$};
\node (oh) at (16.5*\width,\yzero-0.02) {$\Sigma_b^\ast$};
\node (oh) at (17.5*\width,\yzero-0.02) {$\Xi_b^\ast$};
\node (oh) at (18.5*\width,\yzero-0.02) {$\Omega_b^\ast$};

\node (oh) at (-0.6*\width,1.000) {1000}; \draw[black] (-0.01,1.000) -- (0.01,1.000);
\node (oh) at (-0.6*\width,1.250) {1250}; \draw[black] (-0.01,1.250) -- (0.01,1.250);
\node (oh) at (-0.6*\width,1.500) {1500}; \draw[black] (-0.01,1.500) -- (0.01,1.500);
\node (oh) at (-0.6*\width,1.750) {1750}; \draw[black] (-0.01,1.750) -- (0.01,1.750);
\node (oh) at (-0.6*\width,2.000) {2000}; \draw[black] (-0.01,2.000) -- (0.01,2.000);

\node (oh) at (10.4*\width,4.000*\Bscale-\Boffset) {4000}; \draw[black] (11*\width-0.01,4.000*\Bscale-\Boffset) -- (11*\width+0.01,4.000*\Bscale-\Boffset);
\node (oh) at (10.4*\width,4.250*\Bscale-\Boffset) {4250}; \draw[black] (11*\width-0.01,4.250*\Bscale-\Boffset) -- (11*\width+0.01,4.250*\Bscale-\Boffset);
\node (oh) at (10.4*\width,4.500*\Bscale-\Boffset) {4500}; \draw[black] (11*\width-0.01,4.500*\Bscale-\Boffset) -- (11*\width+0.01,4.500*\Bscale-\Boffset);
\node (oh) at (10.4*\width,4.750*\Bscale-\Boffset) {4750}; \draw[black] (11*\width-0.01,4.750*\Bscale-\Boffset) -- (11*\width+0.01,4.750*\Bscale-\Boffset);
\node (oh) at (10.4*\width,5.000*\Bscale-\Boffset) {5000}; \draw[black] (11*\width-0.01,5.000*\Bscale-\Boffset) -- (11*\width+0.01,5.000*\Bscale-\Boffset);
\node (oh) at (10.4*\width,5.250*\Bscale-\Boffset) {5250}; \draw[black] (11*\width-0.01,5.250*\Bscale-\Boffset) -- (11*\width+0.01,5.250*\Bscale-\Boffset);
\node (oh) at (10.4*\width,5.500*\Bscale-\Boffset) {5500}; \draw[black] (11*\width-0.01,5.500*\Bscale-\Boffset) -- (11*\width+0.01,5.500*\Bscale-\Boffset);


\end{tikzpicture}
\caption{\label{fig:comp}(Color online)
Model results and experimental data for the mass differences 
of positive parity heavy baryons and the nucleon. Left panel: charm baryons, right panel:
bottom baryons. The shaded areas are the model results for $x\in[25,30]$ and data are 
indicated by lines and the number (in MeV) is written explicitly. As for ordinary hyperons,
the asterisks denote total spin $j=\fract{3}{2}$.  Note the different scales and off-sets. 
No experimental datum for $\Omega_b^\ast$ is available.} 
\end{figure}
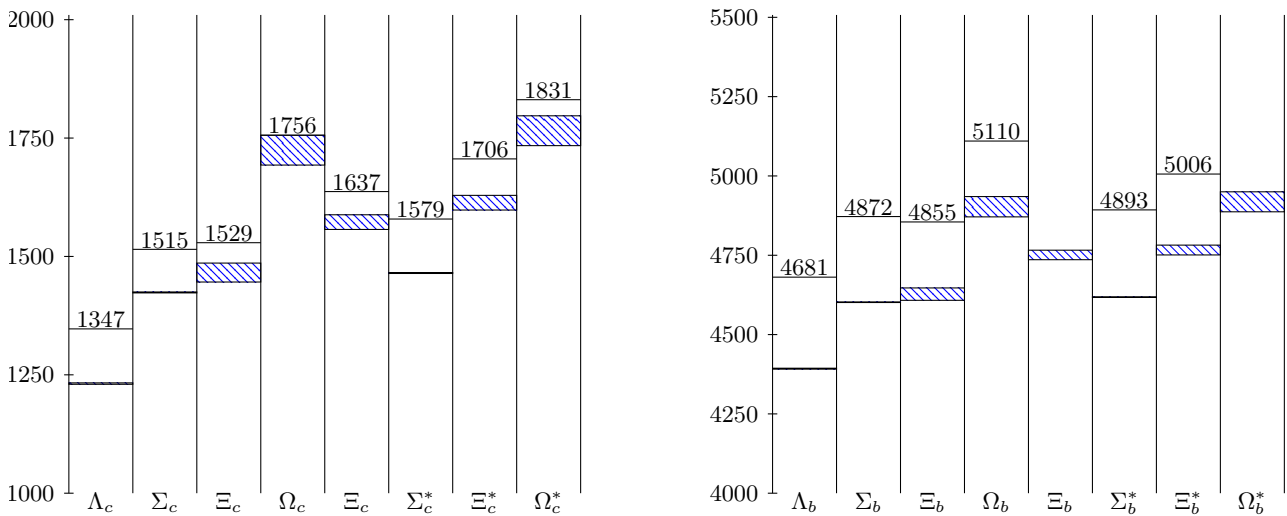

When comparing our model results to data in table \ref{tab:results}
and figure~\ref{fig:comp} we see that the mass differences
within a given heavy quark sector is overestimated. For example 
$M_{\Omega_c}-M_{\Lambda_c}=463{\rm MeV}$ for $x=25$, while the 
empirical value is $409{\rm MeV}$. Further increase of $x$ worsens
the picture. On the other hand, a sizable value ($x\sim30$) for the 
symmetry breaking is required for a good agreement for non-heavy baryons.
Simultaneously the splitting between different sectors is predicted
on the low side. The $\Lambda_c$ and $\Lambda_b$ are about $100{\rm MeV}$ 
and $300{\rm MeV}$ too low, respectively. This is inherited from the heavy 
flavor calculation which overestimates the binding energies in the sense 
that it is too close to the estimate from exact heavy flavor symmetry.
This can also be seen from the parity splitting which is underestimated 
by about $50{\rm MeV}$ (it vanishes in the heavy limit). Together with the 
effect of $SU(3)$ symmetry breaking the overestimated binding combines to 
acceptable agreement for the mass differences between the double strange 
baryons $\Omega_c$ and $\Omega_b$ and the nucleon, at least for $x=30$.
It has been argued~\cite{Harada:1997we} that kinematical corrections due to 
the soliton not being infinitely heavy change the predictions for $\omega_{P,S}$
appropriately. And indeed, replacing the heavy meson masses by the reduced
mass built in conjunction with the classical soliton energy increases
$\omega_P$ by roughly $100{\rm MeV}$ and $\omega_S$ by almost
$200{\rm MeV}$.

For $j=\sfrac{1}{2}$ and positive parity there is an interesting effect in 
the $\Sigma$-$\Xi$ system. The observed mass difference decreases and 
even changes sign when the heaviest flavor turns from strange via charm 
to bottom: $M_{\Xi}-M_{\Sigma}=125,14,-17{\rm MeV}$. 
Partially the model calculation reproduces this effect. For 
example, for $x=25$ the mass differences $101$, $23$ and $6{\rm MeV}$ are
predicted. Since the hyperfine splitting only has a moderate effect, 
the model exhibits a similar scenario for the negative parity channel.
Unfortunately, there are no data to compare with.

Finally we discuss our results for the masses of those strange heavy baryons 
that have previously not been considered in a heavy meson soliton model with 
realistic heavy meson masses: the $\Xi$'s and $\Omega$'s. For the positive parity 
heavy strange baryons we again observe that the mass splittings within a heavy 
multiplet are overestimated. A moderate reduction of the symmetry breaking ratio 
$x$ would be sufficient to match the experimental data. For the negative parity 
$\Xi_c$ with $j=\sfrac{1}{2}$ the too large binding of the $S$-wave reverses this 
picture.  This is not the case for its spin $\sfrac{3}{2}$ counterpart. Interestingly
enough, Ref.~\cite{Agashe:2014kda} assigns the quantum number of this resonance by 
assuming it to join an $SU(4)$ multiplet with the negative parity 
$\Lambda_c(j=\sfrac{3}{2})$. We have argued above that this $\Lambda_c$ is not contained 
in our approach but should be associated with a $D$-wave heavy meson. Thus, as indicated 
in table \ref{tab:results}, it is questionable to identify 
$(I,j,r,p)=(\sfrac{1}{2},\sfrac{3}{2},1,-)$ with $\Xi_c(2815)$. Rather it 
is a prediction for an even heavier resonance like the observed
$\Xi_c(2930)$ or $\Xi_c(2980)$ whose quantum numbers still need to 
be determined~\cite{Agashe:2014kda}.

\section{Conclusion}

We have presented a model calculation for the baryon spectrum that comprises
light and heavy flavors. In particular we have focused on the role of light 
flavor symmetry breaking which is manifested by the strange quark being neither 
light nor heavy. When quantizing the flavor degrees of freedom, the corresponding
deviations from the up-down sector are handled (numerically) exactly.
In the heavy flavor sector the model is inspired by the heavy flavor symmetry, 
with subleading effects arising from finite masses included. The approach 
also includes the hyperfine splitting for the heavy baryons; a moderate effect 
that vanishes in the heavy limit. The model calculation is all-embracing as it 
contains spin $\sfrac{1}{2}$ and $\sfrac{3}{2}$ baryons starting from the 
lightest baryon (nucleon), including hyperons and extending to heavy baryons 
of either parity that have the valence quark content strange-strange-bottom. 
The spectrum is computed from a single mass formula where essentially all parameters 
are determined using data from the baryon number zero sector. We have also calculated 
masses for heavy baryons that are yet to be observed. Though we can only provide an 
estimate for their masses, we find a realistic indication for their positions relative 
to observed baryons.

The overall agreement with data is as expected for chiral soliton model 
estimates. As known from earlier studies, the mass predictions for the heavy
baryons are on the low side when compared to the nucleon. Within a heavy baryon
multiplet the computed mass differences are larger than the experimental data.
This appears to be caused by too strong a remnant of the heavy spin-flavor 
symmetry in the approach. An understanding that goes beyond adopting reduced 
masses in the bound state approach is required. Furthermore the fine-tuning 
of the symmetry breaking ratio $x$ as well as other model parameters that influence 
the soliton properties appears as an obvious endeavor. Actually, a complete 
analysis within a vector meson soliton model (but also a chiral quark model 
as the Nambu-Jona-Lasino model~\cite{Alkofer:1994ph}) shows 
that additional symmetry breaking operators such as 
$\sum_{i=1}^3D_{8i}R_i$~\cite{Park:1991fb} arise in the Hamiltonian, Eq.~(\ref{eq:ham1}). 
Their effects on the heavy baryon spectrum will be reported in a forthcoming paper.

\section*{Acknowledgments}
This work supported in parts by the National Research Foundation NRF,  grant~77454.


\bibliographystyle{apsrev}

\end{document}